# Shoulder Implant X-Ray Manufacturer Classification: Exploring with Vision Transformer

**Simon Meng Zhou**\*, **Marcus Shanglin Mo**\*

School of Computing
Queen's University, 21-25 Union Street, Kingston, K7L2N8
{simon.zhou, shanglin.mo}@queensu.ca

## Abstract

Shoulder replacement surgery, also called total shoulder replacement, is a common and complex surgery in Orthopedics discipline. It involves replacing a dead shoulder joint with an artificial implant. In the market, there are many artificial implant manufacturers and each of them may produce different implants with different structures compares to other providers. The problem arises in the following situation: a patient has some problems with the shoulder implant accessory and the manufacturer of that implant maybe unknown to either the patient or the doctor, therefore, correctly identification of the manufacturer is the key prior to the treatment. In this paper, we will demonstrate different methods for classifying the manufacturer of a shoulder implant.

## 1 Introduction

Shoulder pain is one of the most common issues for most of the people. As the worst case, people may need to take the Total Shoulder Arthroplasty (TSA) to replace the shoulder joint. Today, about 53,000 people in the U.S. have shoulder replacement surgery each year, according to the Agency for Healthcare Research and Quality. As the increasing of amount of people who take the surgery, another issue brought into account: the possible maintenance for the artificial joint after surgery. As we know, there are several different manufacturers producing the artificial shoulder, and each of the manufacturers has several different models to satisfy patients of various shoulder situations.

After years, the implanted shoulders may need to be repaired or replaced. Patient may not remember the model and the manufacturer of their artificial implant. To confirm the model of a shoulder implant, X-ray images taken from the patients and checked by medical experts is a common way. However, with the help of computer vision and deep learning techniques in recent year, we can now use convolutional neural networks to do the classification of a shoulder implant from the X-ray image. Our goal is to prove that machine learning models could perform better for classifying different shoulder implant's manufacturer. We will also compare the performance of classification of traditional non-Neural Network models with some famous Neural Network models. Finally, we explore with an attention-based classification model Vision Transformer.

## 2 Literature Review

### 2.1 Classifying shoulder implants in X-Ray images using deep learning

Gregor Urban, Saman Porhemmat et al. published a paper about classifying shoulder implants in 2020 [1], they use the same data set as we get from the University of California Irvine Machine Learning Repository. In their paper, they represent several non-Neural Network models and seven Neural Network models for evaluating the performance. In our paper, we only selected Random Forest and KNN (K-Nearest Neighbor) for non-Neural Network baseline models and leveraged VGG-16 and ResNet-50 as a comparison to the result in Urban et al.'s paper [1]. Additionally, other new models that not from that paper are reported as well. We also explore with different data augmentation techniques to boost the accuracy and make the Neural Network models more robust, thus prevent overfitting.

---

\*Co-Author

## 2.2 Automatic Detection and Segmentation of Shoulder Implants in X-Ray Images

This paper is Maya Belen Cervantes Gautschi Stark's master thesis that submitted in 2018 [2]. Unfortunately, this paper does not focus on classifying which manufacturer it belongs to for a given shoulder implant X-ray image. The model introduced in this paper, based on Hough Circle Detection, only serves for determine whether there is a shoulder implant or not by looking at the X-ray image and segment the actual implant out from the image.

Although there is another study utilizes deep learning technique to improve the fracture detection in X-ray images for many of the body parts, the determination of manufacturer of a particular shoulder implant is a new task in the field medical image classification. In particularly, we are the first group leveraging the attention-based Neural Network model for this specific task.

## 3 Data

### 3.1 Original Data

The shoulder implants X-ray images were obtained for a dataset that available from the University of California Irvine Machine Learning Repository. This dataset contains a total of 597 X-Ray images from the following manufacturers: 83 images belong to Cofield, 294 images belong to Depuy, 71 images belong to Tornier and 149 images belong to Zimmer. Notice that the dataset has the class imbalance problem. A general and naïve classifier without any tunning could classify every image to the dominant class "Depuy" with a 49.17% accuracy. Therefore, to boost the accuracy and allow the model to classify other class other than "Depuy", some data preprocessing techniques are introduced in this paper. Figure A. simply shows four different manufacturer's shoulder implants.

One of the preprocessing methods is to standardize the image. We applied normalization of each pixel in the image so that every image is in the shape of 256*256*3 and each pixel value inside the image has range of [0, 1].

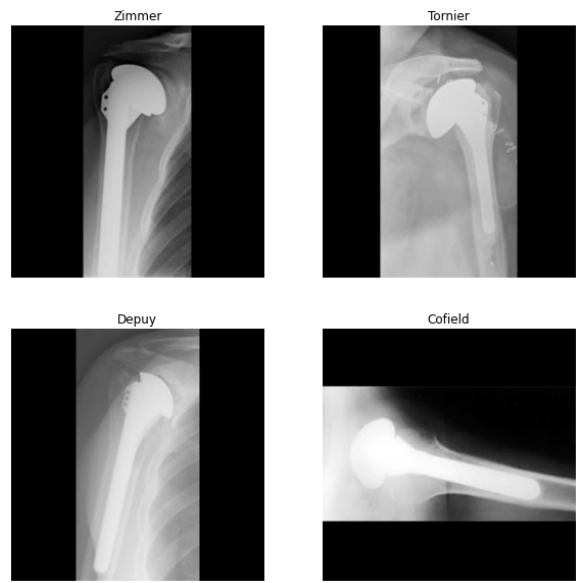

Figure A. A simple visualization of different manufacturer's implant

### 3.2 Data Augmentation

Deep learning models often requires a huge amount of training data. However, there are only 597 images used in this task, which is an insufficient amount of training an accurate model for classifying each class. Therefore, the data augmentation technique is used to increase the number of training data. Data augmentation is the process of applying different type of transformations to each image in the dataset to enlarge the dataset. In this paper, random scaling, cropping, translation, rotation, distortion, horizontal and vertical flip, and addition of gaussian noise is considered as the augmentation method. Therefore, each image has another of eight images corresponding with it after data augmentation, form a total of nine images (including the original one). A simple image after data augmentation is shown in Figure C.

For the sake of simulating the real-life case, we randomly select 25% of original image, about 150 images as the final test dataset only when the model is **trained** on the **augmented data**. These 150 images are purely test data, they will not go under data augmentation and will not be used until we reach the final evaluation stage. For **training** on the **original data**, a simple train test split technique is applied.

After data augmentation, we have a total of 4023 images in training data and 150 images in test data. However, due to RAM limitation in Google Colab, the large number of images will cause RAM crash for most of the Neural Network models, even though we set a very small batch size in the training process. As the result, we further reduce the size of data augmentation by only selecting random rotation, scaling, cropping and horizontal flip as the augmentation techniques. By doing so, the new augmented images reduce to 2235 images for training and 150 images for testing. The visualization after reducing the augmentation size is shown in Figure B. Images were augmented via *imgaug* library with Python 3.6.

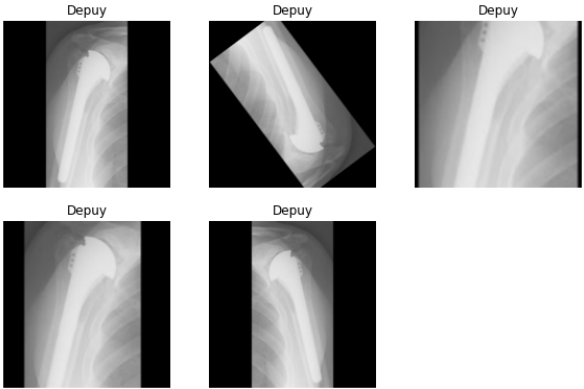

Figure B. A simple visualization of data augmentation after reducing its size

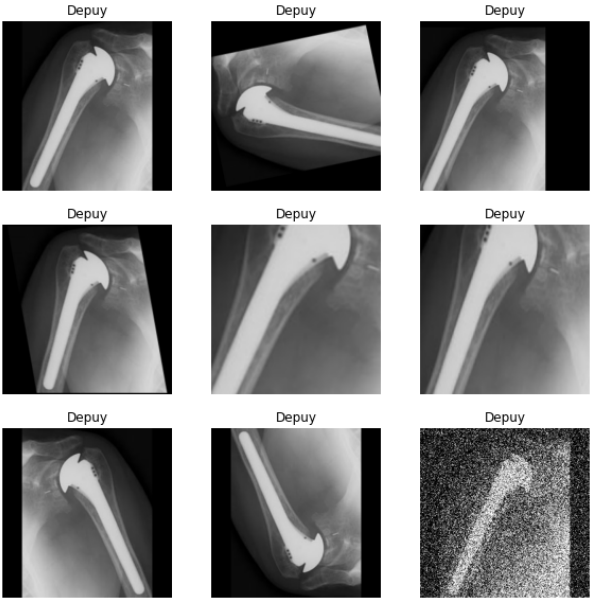

Figure C. A simple visualization of an image after data augmentation

### 3.3 Data Augmentation Pipeline

In this section, we present our data augmentation pipeline. The pipeline can be modularized into three sections:

**Split Data** INPUT: a path directory that contains all images. OUTPUT: a list of images for augmentation and a list of images for only testing.

**Data Augmentation** INPUT: a list of images for augmentation. OUTPUT: perform data augmentation, each image in the list produces other four augmented images and converted to numpy array for fitting in the model.

**Test Data Processing** INPUT: a list of images for testing. OUTPUT: the numpy array containing the test data.

## 4 Experiments

The first couple of attempts were focus on the traditional machine learning algorithms including Random Forest and K-Nearest Neighbor for computing the baseline score of this task. Deep methods were also introduced though VGG-16, ResNet-50 and Inception-V3 architecture. Lastly, we utilized Vision Transformer for classification the four manufacturers. These methods were chosen to illustrate the classification performance in our paper. The section below discussed the traditional machine learning algorithms first and followed by the deep methods.

### 4.1 Traditional Machine Learning Models

Similar to Urban et al.'s [1] work, Random Forest and K-Nearest Neighbor algorithm was applied on this dataset with and without using data augmentation. The Random Forest model is tunned via Randomized Search Cross-Validation, whereas KNN model is not tunned due to the limitation of RAM in Google Colab. These models operate directly on the pixel's value of each image and the score of each model is used as a benchmark for comparing the performance of the Deep models later. The traditional machine learning models were directly implemented via *sklean* library with Python 3.6.

#### 4.1.1 Random Forest:

Starting with the Random Forest classifier with entropy as the loss function for this task. The hyperparameters used in the model are the number of trees (n_estimator); the number of max features per leaf (max_features) and the minimum data samples per leaf (min_samples_leaf). Randomized Search Cross-Validation (Randomized SearchCV) been used in this model to find the best parameter set and could reduce the computational time and RAM usage. The best parameters set returned from the Randomized SearchCV is 500 decision trees, minimum 1 data sample per leaf and maximum of 3 features per leaf. The model reshapes the data into two-dimensional array by multiplying image's height, width, and the number of channels. The performance of Random Forest with and without data augmentation are listed in Table 1.

#### 4.1.2 K-Nearest Neighbor

A K-Nearest Neighbor classifier was also studied in this paper. The classifier used the Euclidean distance as the measurement metric and with $k = 30$. K-Nearest Neighbors algorithm determines the label of a new given sample by computing the Euclidean distance and find the k nearest neighbor to that new sample. Then, a majority voting technique is used to determine the label of the new sample. A relatively large $k$ was used in this task because there are only four classes and some images in different classes look quite same. The performance of K-Nearest Neighbor with and without data augmentation are listed in Table 1.

### 4.2 Deep Learning Models

In the past few years, deep learning has a significant impact in the field of image classification. Deep learning models achieves a very high accuracy on some complex problems, which is a very far beyond what a traditional machine learning models could archive in the same type of problems. Specifically, Convolutional Neural Networks (CNN), a very powerful Neural Network structure, was able to detect both low and high dimensional features through the convolutional filters [3,4]. The filter slides through the image and extract the local feature each time, util the whole image has been studied. CNN also could reduce its number of parameters by introducing the pooling layer. This layer typically finds the maximum value or compute the average value for each filter. A simple Convolutional Neural Network is shown is Figure D.

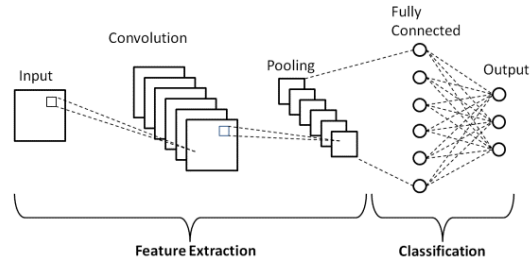

Figure D. A simple CNN structure [9]

In this paper, three popular convolution based Neural Network architectures were studied, namely VGG-16, ResNet-50 and InceptionV3. These three architectures were proven to work on classification task with success based on [1,4], [5] and [6]. VGG-16 and InceptionV3 were implemented by using Keras Applications API while ResNet-50 was implemented by both hand and API. Both ways of implementation are based on Python 3.6 in Google Colab with TensorFlow and Keras package. The hyperparameter tuning process of models will not be explicitly explained, the process is done during the training process. The optimizer of these methods is either Adam or Stochastic Gradient Descant with $lr = 0.001$. The score of deep learning methods is also in Table 1.

### 4.2.1 VGG-16 Architecture

VGG-16 was selected as the first deep learning model for this task. It is a very popular CNN architecture that is always used for image classification task, it also achieves a high accuracy score very often. Although VGG-16 is a very powerful architecture, it requires around 144 million parameters, and it is very time consuming to train on the laptop. The architecture composed of many 3x3 convolutional filters with a stride of 1 and max pooling layers of 2x2 window with stride of 2. After the convolutional operation is done, the network adds several fully connected layers with 2048 units and finally with a 1000 dimension of output for 1000 classes in ImageNet classification task [4]. In this paper, VGG-16 only trained and tested without data augmentation through transfer learning and from scratch. Table 1. shows that accuracy score of this model and Figure E. shows the VGG-16 architecture:

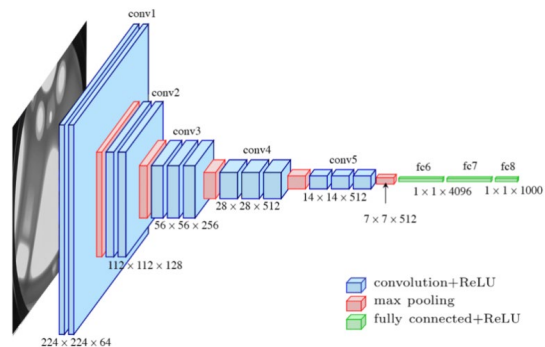

Figure E. VGG-16 Architecture [10]

### 4.2.2 ResNet-50

ResNet-50, also known as 50-layer Residual Network, is another popular Convolutional Neural Network that proven to be powerful for classification problems. It is also a powerful model to perform transfer learning. There are different variations of Residual Networks, but ResNet-50 provides a sufficient amount of layers (depth) in the classification problem. The model also addresses the problem that other general deep Neural Networks may have, which is the degradation problem. This problem happens when the network is too deep to maintain the high accuracy, once the accuracy becomes saturated, continue training may cause it to degrade dramatically [5]. The Residual Net model contains the skip connection feature, which will be adding some of the inputs of stacked convolutional layers to the output directly. This will reduce the chance of gradient vanishing. The identity map in the model ensures that the network will learn from identity mapping to prevent the accuracy dropping because of saturation.

In this paper, the ResNet-50 model with transfer learning was studied, the cross-validation technique also used to increase the accuracy of the model. A simple Residual Network architecture is shown in Figure F. and the relevant score is in Table 1.

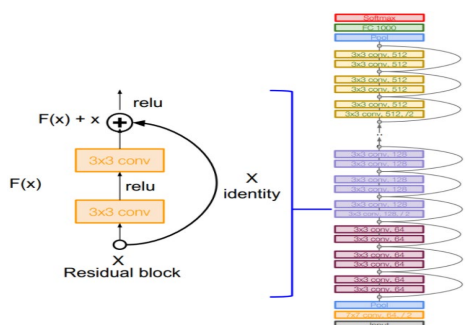

Figure F. Residual Network Architecture [11]

### 4.2.3 InceptionV3

InceptionV3 is also a powerful network for image classification and recognition, and it is invented by Google Research in 2015. Unlike the two models discussed above, the general Inception model replaces the determination of convolution filter size with several common filter sizes (1x1, 3x3, 5x5), then concatenate the result from those filter size as the input of pooling and fully connected layer. As for the InceptionV3 model [7], a new way of representing convolution filter is discussed. The model introduces a new concept called "Factorization into small convolutions", which is simply replace a $n \times n$ filter by a $n \times 1$ and a $1 \times n$ filter. This method will reduce the number of parameters and prevent overfitting.

In this paper, a InceptionV3 model was implemented via Transfer Learning and performed cross-validation on this model. A InceptionV3 architecture is shown on Figure G and the relevant score is in Table 1.

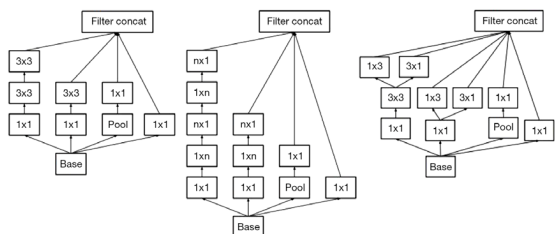

Figure G. InceptionV3 Architecture [12]

### 4.2.4 Vision Transformer

Transformer is a popular and powerful model in the field of Nature Language Processing. The basic idea behind transformer is self-attention [7], where each word is connected to every other word in an NLP model. The attention method allows the model to focus on the "important" feature of the next input. Similarly, the Vision Transformer (ViT) [8] uses the same idea of Transformer in NLP. The idea behind Vision Transformer is using the encoding part to perform classification.

The input image is separated by many small patches and then flattened into a linear shape. Each patches of image have been converted to a numpy array as the input of Transformer encoder, and a learnable class token are also pass into the encoder for classification propose.

In this paper, the transfer learning approach of Vision Transformer and train from scratch approach were both applied in the task. To demonstrate the result of splitting image into patches, a random shoulder implant image was chosen and performed patching on it. Figure H. shows splitting an image into several 20 × 20 patches. Figure I. shows the network structure for Vision Transformer and the relevant score is in Table 1.

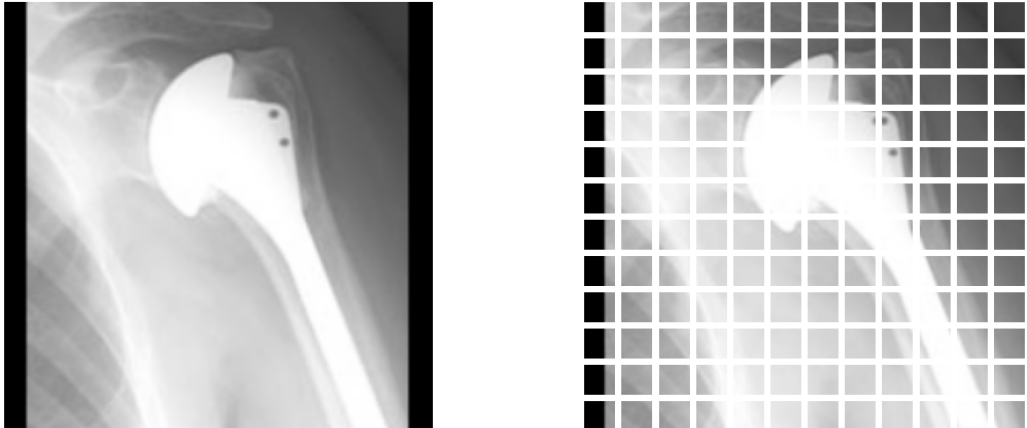

Figure H. Splitting image into 12 20 × 20 patches

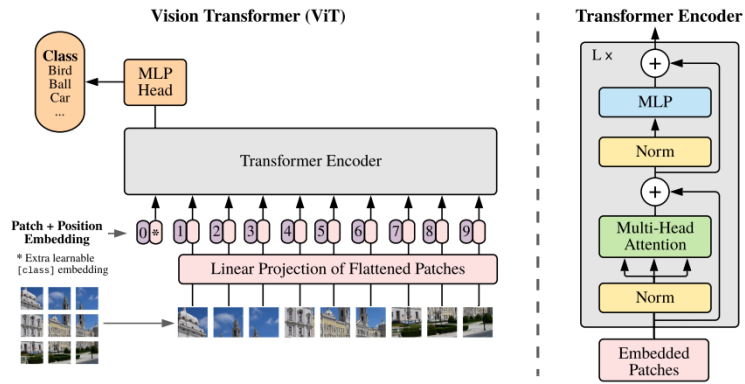

Figure I. Vision Transformer structure [8]

### 4.2.5 Transfer Learning

Usually transfer learning is a good approach to the task where the data is limited. In this task, there are only 600 images and 2235 images before and after data augmentation, which in this case, both are insufficient for a general image classification task. Therefore, transfer learning technique been applied in this paper. All the models introduced above are used for transfer learning with ImageNet's weights. The convolutional layers in the network will serve to extract the local feature of any input images, so the weights for those layers are frozen so that are not trainable. The model will only train on the very last couple of custom fully connected layers for the shoulder implants task.

## 5 Results and Analysis

As discussed above, the classification performance of each model introduced in this paper were collected and shown in Table 1. Only Random Forest, ResNet-50 and InceptionV3 used cross-validation technique. Other models were trained in a single fold due to the RAM limitation in Google Colab. In this task, the accuracy and precision score are selected to be the one to evaluate the performance of models.

### 5.1 Classification Accuracy

A naïve way to represents the performance is looking at the classification accuracy. The accuracy score represents the ratio of number of samples that classified correctly by the total number of samples.
The accuracy is computed as the following:

$$\text{Classification accuracy} = \frac{correct\ number\ of\ prediction}{total\ number\ of\ prediction}$$

The accuracy score for each model is recorded in Table 1. By observing the table, deep models, especially ResNet-50, achieves the highest accuracy score of 77% on the 150 images test dataset. The baseline accuracy score is 49.17% from traditional machine learning algorithms without data augmentation. Since a naïve classifier could achieves around 50% of accuracy by simply guessing the dominant class "Depuy", the accuracy score may not be the perfect metric for evaluating the performance in this task.

### 5.2 Classification Precision

To find the accuracy within each class, a precision score is calculated. The precision score represents how much the model correctly predicted in each class in this scenario. The precision is calculated as the following:

$$\text{Classification precision} = \frac{Total\ number\ of\ correct\ prediction\ in\ each\ class}{total\ number\ of\ samples\ in\ each\ class}$$

By using this metric, the model will tell us the classification performance per class, not with the whole dataset. The baseline precision score is 50% or less, obtained by the traditional machine learning algorithms. The best precision score achieved is 81% by ResNet-50. The precision is more reasonable than the accuracy for shoulder implants manufacturer classification task.

### 5.3 Experimental Results

#### 5.3.1 Table of scores:

| Model | Augmented | VA | TA | TP |
|---|---|---|---|---|
| Random Forest | No | 0.53 | N/A | 0.50 |
| Random Forest | Yes | N/A | 0.48 | 0.40 |
| K-Nearest Neighbor | No | 0.53 | N/A | 0.50 |
| K-Nearest Neighbor | Yes | N/A | 0.41 | 0.40 |
| VGG-16-scratch | No | 0.45 | N/A | 0.38 |
| VGG-16-TL | No | 0.49 | 0.49 | 0.49 |
| ResNet-50-TL | No | 0.49 | 0.49 | 0.49 |
| ResNet-50-TL-cv | Yes | 0.88 | **0.77** | **0.81** |
| InceptionV3 | No | 0.66 | N/A | 0.61 |
| InceptionV3-TL-cv | Yes | 0.77 | 0.55 | 0.51 |
| Vision Transformer | Yes | 0.50 | 0.43 | 0.34 |
| Vision Transformer-TL | Yes | 0.88 | **0.68** | **0.71** |

Table 1. scores for all models implemented VA: validation accuracy, TA: test accuracy, TP: Test Precision, TL: applied Transfer Learning, CV: applied cross-validation

#### 5.3.2 Confusion Matrix:

The Figure J. shows the confusion matrix of ResNet-50 with transfer learning and 10-fold cross validation on the 150 images test dataset.

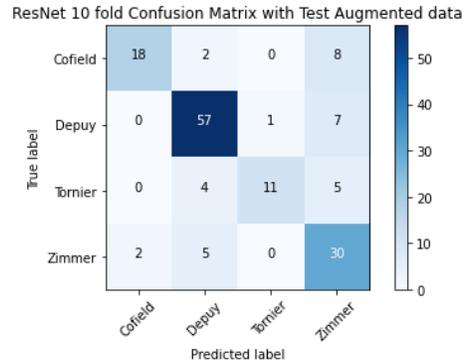
Figure J. ResNet50-TL-CV confusion matrix

The figure K. shows the confusion matrix of Vision Transformer with transfer learning.

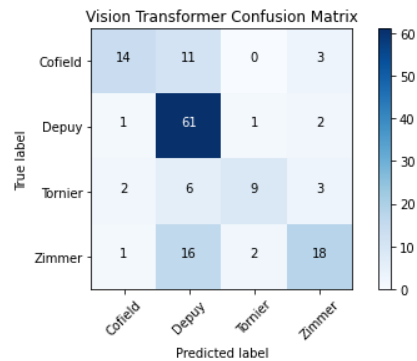
Figure K. ViT-TL confusion matrix

Vision Transformer usually performs better when there is a large amount of data provided. Given only around 2000 images after augmentation, the performance is decent under this scenario. In fact, the precision score of Vision Transformer is better than all other models except ResNet-50.

5.4 Analysis

Through this paper, deep learning method is proven to be successful in the classification of shoulder implant's manufacturers. The ResNet-50 with transfer learning and cross-validation that trained on the augmented data demonstrates the best precision score overall. Convolutional Neural Network works excellent on the dataset with images. Its unique two-dimensional filters were used to catch the local feature of each image to be able to distinguish between different types of shoulder implants.

Transfer learning also helps with the increment of accuracy and precision score in deep learning models. With transfer learning, the validation accuracy score of VGG-16 model increases about 4%; InceptionV3 model increases about 11% and Vision Transformation increases 28%. Moreover, the usage of cross-validation is also important for the model. The ResNet-50 model after 10-fold cross validation achieves 77% in the test accuracy. The cross-validation method used in this paper is Stratified K-Fold cross-validation. This method is suitable when the data provided are suffering class imbalance problem. In each split, the Stratified K-Fold tries to split the data into the same ratio of each class in the original dataset (the ratio of each class in training and validation dataset are the same as the class ratio in the original dataset). It ensures that in the extreme situation, the model could learn for at least one image per class.

Another significant finding is the power of data augmentation. Data augmentation provides additional data for training, it could enlarge the variation of the existing data, lower the chance of overfitting and boosting the accuracy of models. The ResNet-50 model almost increases 40% of its accuracy by utilizing the data augmentation technique. This technique brings the most significant impact for Convolution based Neural Network. For traditional machine learning algorithms, increasing the amount of data does not increase the accuracy, this is because the traditional algorithms could directly reshape the image into a one-dimensional array no matter what, thus lose the relevant position information. Adding more data will only increase the chance of overfitting.

The Vision Transformer model also works fine, it may require more data to classifies the right class. The self-attention mechanism is very power not only in the field of NLP, but also in Computer Vision. Splitting the image into many patches helps the model to learn the image better, when sending these patches into transformer encoder, the self-attention mechanism is applied. It will look for the most significant feature for each class and predict a new input image based on the significant part.

It is risky to pull out 150 images as test data when using augmentation to train model, because we have a few images in each class in the original dataset. Pulling out 150 images for testing will directly affect the model performance. However, to make the model more generalized and closer to the real-life scenario, this decision has been made eventually.

# 6 Conclusion

Deep learning methods are effective in the shoulder implants manufacturer classification task through X-ray imaging. Data augmentation, transfer learning and cross-validation are used to improve the model performance. With more data and perhaps more manufacturers, the existing deep learning model would perform better, and a tool could be created to assist doctor to classify shoulder implants in the real life. Doctors could know which shoulder implant belongs to which manufacturer in a very short period of time by using the deep learning model and perform correct replacement or repair in time.

Feature improvements of this work would be considered other deep learning models. Improving the performance of Vision Transformer is also a challenge task. Someone may study how the image segmentation technique could connected to the image classification task if possible. Also, an assistant tool of classifying shoulder implants should be considered. The tool could not only be used in the PC computer, but also be used in the mobile phone.

# 7 Team Contribution

Simon Meng Zhou is charged for all the deep learning models construction, perform the transfer learning, cross-validation, and fine tune the model. He also studied the Vision Transformer that present in this paper.

Marcus Shanglin Mo is charged for all the traditional machine learning algorithms, perform hyperparameter tuning via Randomized Search. He also charged for presentation slides and formatting.

Data augmentation process is done by both Meng and Shanglin.

# 8 Acknowledgements

We would like to acknowledge our professor of this course Dr. Steven H. Ding for his teaching in this semester and continuously providing us the background knowledge of different algorithms to properly utilize this is task.

# 9 Other Materials

The code of this paper could be found at https://github.com/simonZhou86/Advanced-Data-Analytics-Project

The original dataset is also on the link above, under Data set folder, or you can also download at https://archive.ics.uci.edu/ml/datasets/Shoulder+Implant+X-Ray+Manufacturer+Classification#